%% file: main.tex
\documentclass[sigconf,nonacm]{acmart}
\input{dlabmacros.tex}



\usepackage{subcaption}

\usepackage{tikz}
\usetikzlibrary{positioning, calc, shapes.geometric, shapes, shapes.multipart, arrows.meta, arrows, decorations.markings, external, trees}
\tikzset{
node distance=0.5cm, 
}
\tikzstyle{Arrow} = [
	thick, 
	decoration={
		markings,
		mark=at position 1 with {
			\arrow[thick]{latex}
			}
		}, 
	shorten >= 0.5pt, preaction = {decorate},
	]

\AtBeginDocument{%
  \providecommand\BibTeX{{
    \normalfont B\kern-0.5em{\scshape i\kern-0.25em b}\kern-0.8em\TeX}}}

\copyrightyear{2023}
\acmYear{2023}
\setcopyright{acmlicensed}\acmConference[WWW '23]{Proceedings of the ACM Web Conference 2023}{May 1--5, 2023}{Austin, TX, USA}
\acmBooktitle{Proceedings of the ACM Web Conference 2023 (WWW '23), May 1--5, 2023, Austin, TX, USA}
\acmPrice{15.00}
\acmDOI{10.1145/3543507.3583275}
\acmISBN{978-1-4503-9416-1/23/04}

\begin{document}

\title[Automated Content Moderation Increases Adherence to Community Guidelines]{Automated Content Moderation Increases Adherence to Community Guidelines}


\author{Manoel Horta Ribeiro}
\affiliation{%
  \institution{EPFL}
  \country{Lausanne, Switzerland}
}
\email{manoel.hortaribeiro@epfl.ch}

\author{Justin Cheng}
\affiliation{%
  \institution{Meta}
    \country{Menlo Park, CA, USA}
}
\email{jcheng@meta.com}

\author{Robert West}
\affiliation{%
  \institution{EPFL}
  \country{Lausanne, Switzerland}
}
\email{robert.west@epfl.ch}

\renewcommand{\shortauthors}{Horta Ribeiro et al.}

\begin{abstract}
Online social media platforms use automated moderation systems to remove or reduce the visibility of rule-breaking content.
While previous work has documented the importance of manual content moderation, the effects of automated content moderation remain largely unknown.
Here, in a large study of Facebook comments ($n=412$M), we used a fuzzy regression discontinuity design to measure the impact of automated content moderation on subsequent rule-breaking behavior (number of comments hidden/deleted) and engagement (number of additional comments posted).
We found that comment deletion decreased subsequent rule-breaking behavior in shorter threads (20 or fewer comments), even among other participants, suggesting that the intervention prevented conversations from derailing.
Further, the effect of deletion on the affected user's subsequent rule-breaking behavior was longer-lived than its effect on reducing commenting in general, suggesting that users were deterred from rule-breaking but not from commenting.
In contrast, hiding (rather than deleting) content had small and statistically insignificant effects.
Our results suggest that automated content moderation increases adherence to community guidelines.
\end{abstract}

\ccsdesc[500]{Human-centered computing~Empirical studies in collaborative and social computing}
\keywords{content moderation, online platforms, community guidelines}


\maketitle


\begin{figure}
\centering
\includegraphics[width=0.9\linewidth]{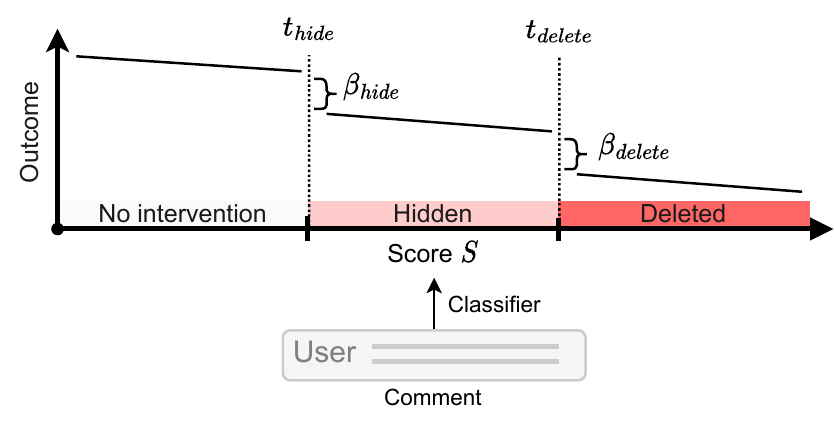}
\caption{
Comments posted on Facebook are scored by classifiers that measure adherence to community standards. 
When $S$ crosses specific thresholds ($t_\text{hide}$ and $t_\text{delete}$ in the figure), different interventions are applied.
Though comments around each threshold are similar, they receive different interventions, \eg, a comment with score $t_\text{delete} - \epsilon$ is hidden, while a comment with score $t_\text{delete} + \epsilon$ is deleted.
Exploiting this fact, this work measures the impact of these interventions on user behavior outcomes by studying the discontinuities ($\beta_{hide}$ and $\beta_\text{delete}$) around the thresholds $t_\text{hide}$ and $t_\text{delete}$.
}
\label{fig:lead}
\end{figure}

\section{Introduction}

Many online platforms enforce community guidelines using automated content moderation systems that detect and intervene when rule breaking occurs, 
\ie, when user behavior violates community guidelines~\cite{transparency_fb, transparency_yt, transparency_tiktok, transparency_twitter}.
These systems prevent harm by removing or reducing the visibility of rule-breaking content~\cite{grimmelmann2015virtues}, \eg, by reducing the number of people who see such content~\cite{gelber2016evidencing}.
However, content removal or visibility reduction may also affect \textit{on-platform} user behavior~\cite{kraut2012building, srinivasan2019content}.
Moderation interventions may increase compliance with community guidelines, \eg, as deleted comments may prevent a conversation from derailing~\cite{zhang2018conversations}, or, reversely, backfire and increase rule breaking, \eg, because sanctioned users perceive the decision as unfair~\cite{chang2019trajectories}.

\begin{figure*}
\centering
\includegraphics[width=\linewidth]{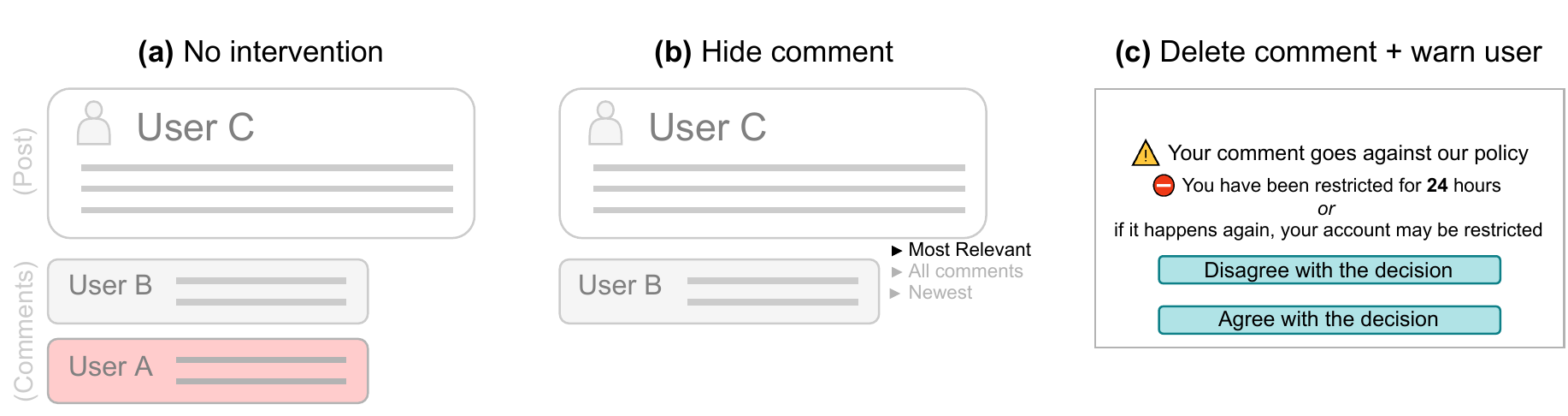}
\caption{
Moderation interventions -- We depict a hypothetical scenario where \textit{User A} writes a rule-breaking comment (in red) on a post by \textit{User C} (white) that has received a comment by \textit{User B}, (grey).
Depending on the score a comment receives, (a) no intervention may be applied, in which case the comment is posted, (b) the comment may be \emph{hidden}, and it will only be visible if a viewer changes the default comment ranking setting to show all comments, or (c) the comment may be \emph{deleted} and the user who posted the comment warned (an additional sanction may be applied depending on their previous rule-breaking behavior). 
}
\label{fig:interventions}
\end{figure*}

Understanding the causal effect of automated content moderation practices on user behavior is vital for evaluating these systems' effectiveness and can inform their design and use.
However, measuring the causal effect of content moderation is difficult because of the ethical and technical challenges in using randomized experiments (e.g., A/B testing) to study content moderation practices~\cite{srinivasan2019content}.
Allowing some users not to be moderated implies not removing content that may harm others, and malicious actors could exploit the randomization of potential experiments to post harmful content.

Previous work has extensively documented the role of ``manual'' content moderation in online communities~\cite{kraut2012building, seering2020reconsidering}, \ie, where volunteer moderators find and remove content that breaches community guidelines.
Some research has sought to estimate the effect of such manual content moderation on online communities, finding that it positively impacts user behavior~\cite{seering2017shaping,srinivasan2019content}.
Nevertheless, these effects may not generalize to \emph{automated}, platform-level content moderation.
Further, research on content moderation has been typically descriptive~\cite{cheriyan2017norm, fiesler2018reddit,chandrasekharan2018internet,gilbert2020run, butler2008don, jhaver2019human} rather than causal, and the quasi-experimental designs used in previous work are not readily adapted to an automated setting.
For example, some approaches that rely on the randomness in time it takes for human moderators to intervene upon rule-breaking content to estimate the effect of content moderation \cite{srinivasan2019content} do not work for automated systems, in which moderation occurs immediately after content is created.

\xhdr{Present work} 
We study the effect of automatically enforcing community guidelines for violence and incitement for Facebook comments on user behavior with a quasi-experimental approach illustrated in \Figref{fig:lead}.
We examined subsequent rule-breaking behavior and commenting activity among users whose comments were moderated (user-level scenario) and among users in threads where these comments were posted (thread-level scenario).
Analyzing over $412$M comments,  we measured the effect of two different interventions (hiding and deletion; see \Figref{fig:interventions}) on outcomes capturing commenting activity and rule\hyp breaking behavior (see \Secref{sec:methods}).
Specifically, we estimated the causal effect of content moderation using a fuzzy regression discontinuity design~\cite{imbens2008regression}, an approach that capitalizes on two design choices commonly used in automated content moderation systems~\cite{teblunthuis2021effects, Coral}.
First, on many platforms, machine learning models that predict whether content breaches community guidelines assign a score of $S$ to each piece of content, reflecting the likelihood of that content breaking community guidelines.
Second, when these assigned scores are sufficiently high, platforms may \emph{automatically} enforce community guidelines.
In other words, if a score exceeds a predetermined threshold $t$ and certain other conditions are met, content may be hidden or deleted immediately.%
\footnote{The regression discontinuity is \textit{fuzzy} because there is a chance that units (\ie, comments) below the threshold may be treated (\ie, intervened upon) and units above it may not be treated (\ie, not intervened upon).
This can happen because other mechanisms can trigger or prevent interventions on comments with scores below or above the thresholds (\eg, users may manually report comments under the threshold; other systems may exclude some comments from intervention).}
Our approach allows us to mimic a randomized control trial around the threshold $t$ since content with a score right above the threshold (\ie, $S = t + \epsilon$) is similar to content with a score right below (\ie, $S = t - \epsilon$), but only the former is automatically intervened upon by the content moderation system (which may, \eg, delete it). 

\xhdr{Results}
Overall, deleting comments reduced rule-breaking behavior in the thread where the comment was originally posted.
Deleting comments also reduced rule-breaking among users whose comments were deleted, \ie, other comments in the thread or that the user subsequently posted were hidden and deleted less often after the intervention.
At the thread level, deleting rule-breaking comments significantly decreased rule-breaking behavior in threads with 20 or fewer comments before the intervention, even among other participants in the thread.
This effect was statistically insignificant for threads with more than 20 comments.
Deletion at the user level led to a decrease in subsequent rule breaking and posting activity. But while the decrease in rule breaking persisted with time, the decrease in posting activity waned. In other words, users gradually returned to making posts or comments at a rate similar to before their comments were deleted but were less likely to post comments that would subsequently be hidden or deleted.
Hiding (rather than deleting) content had small and statistically insignificant effects on subsequent user activity and rule-breaking behavior at both the user and thread levels.

\xhdr{Implications}
Deletions of rule-breaking content by automated content moderation, as currently applied on Facebook, decrease the subsequent creation of content that goes against community guidelines.
Our results suggest two ways that this may happen.
First, users whose comments are deleted are less likely to produce subsequent rule-breaking content.
Second, other users are also less likely to create rule-breaking comments in the thread where the content was deleted.
Building on previous work that found that ``manual'' content moderation~\cite{seering2017shaping, srinivasan2019content} can prevent rule-breaking behavior, here we show that these effects generalize to automated systems responsible for a substantial fraction of moderation interventions carried out by major social networking platforms~\cite{transparency_fb, transparency_yt, transparency_tiktok, transparency_twitter}.
Though our results are limited in that we can only measure the effect of content moderation interventions triggered by classifiers at the thresholds at which they are applied, this study may clarify their present impact on online platforms such as Facebook.
And while automated content moderation systems are typically assessed using precision and recall, this work shows how they may also be evaluated in terms of their effects on subsequent user behavior in an observational manner that does not require experimentation.

\section{Related Work}
Our investigation builds on two bodies of work -- research on anti-social behavior and content moderation.

\xhdr{Anti-social behavior}
Anti-social behavior has been present on social media since its early days~\cite{dibbell1994rape}. 
Given its detrimental effect on people's lives~\cite{akbulut2010cyberbullying,wiener1998negligent,gelber2016evidencing}, a vast body of research has characterized it across a variety of platforms, languages, and contexts~\cite{cheng2017anyone,wulczyn2017ex,thomas2021sok}.

One line of work has used the growing capabilities of machine learning models to detect cyberbullying~\cite{rafiq2015careful}, hate speech~\cite{kiela2020hateful}, trolling~\cite{kumar2014accurately}, and online harassment~\cite{stoop2019detecting}.
Many commonly-used classifiers generate scores that are subsequently used to determine when intervention is appropriate.
For example, Google Perspective's flagship classifier~\cite{perspective} outputs a ``toxicity'' score for short texts that reflects ``rude, disrespectful or unreasonable comments that are likely to make someone leave a discussion.''
This score has been used to proactively intervene upon potentially rule-breaking content, for instance, on Coral, an open-source commenting platform used in over 100 newsrooms, including the Washington Post and Der Spieger~\cite{Coral}.
But while classifiers for detecting undesirable behavior exist, less research focuses on understanding their impact. For instance, previous work has highlighted how such systems may struggle with context and differences in dialects~\cite{sap2019risk,ribeiro2018characterizing}.

Another line of research relevant to the present work has examined whether anti-social behavior is ``contagious,'' \ie, studying a user's likelihood to produce trolling or uncivil content after being exposed to similar content.
Findings have been mixed: some papers have found that rule-breaking behavior can spread from comment to comment~\cite{cheng2017anyone, kim2021distorting}, while others have found null results~\cite{han2015playing,rosner2016dangerous}.

\xhdr{Content moderation}
A majority of prior work on content moderation has examined how it occurs at the community level, where members of the community enact sanctions (\eg, elected ``administrators'' on Wikipedia~\cite{butler2008don}), rather than at the platform level, where centralized agents shape moderation decisions~\cite{seering2020reconsidering}.
This prior work has focused on describing content moderation practices and governance systems in online communities either quantitatively~\cite{cheriyan2017norm, fiesler2018reddit, chandrasekharan2018internet} or qualitatively~\cite{gilbert2020run, butler2008don, jhaver2019human}.
For instance, research characterizing rule-breaking behavior on Reddit found that some norms were universal while others were unique to specific subreddits \cite{chandrasekharan2018internet}.

More aligned with the work at hand is research evaluating the effects of content moderation at both the community and platform levels.
At the community level, past research has measured the effectiveness of removing content on Reddit~\cite{srinivasan2019content} or providing explanations for removals~\cite{jhaver2019does}, finding that both reduced subsequent rule-breaking behavior.
Research has also explored how proactive moderation tools like chat modes on Twitch~\cite{seering2017shaping} or post approvals on Facebook groups~\cite{ribeiro2022post} can prevent anti-social behavior.
Yet other work studied the effect of algorithmic flagging on Wikipedia using a sharp regression discontinuity design, finding that the system leads to more fair outcomes on the platform~\cite{teblunthuis2021effects}, although effects are heterogeneous across language editions with different characteristics~\cite{wang2022ml}.
At the platform level, previous work on the effects of content moderation has analyzed soft-moderation strategies, \eg, flagging news as misinformation~\cite{zannettou2021won,mena2020cleaning,seo2019trust}, and the effect of deplatforming users and communities from mainstream platforms~\cite{ jhaver2021evaluating,horta2021platform,chandrasekharan2017you}.

\xhdr{Relationship between prior and present work}
Prior work has studied online moderation, often in a descriptive fashion~\cite{cheriyan2017norm, fiesler2018reddit,chandrasekharan2018internet,gilbert2020run, butler2008don, jhaver2019human} or manual, community-oriented contexts~\cite{seering2017shaping,ribeiro2022post,jhaver2019does,srinivasan2019content}.
Yet less is known about the impact of large-scale platform-level automated online moderation (a substantial fraction of moderation interventions carried online \cite{transparency_fb, transparency_yt, transparency_tiktok, transparency_twitter}).
Therefore, the results provided here advance the understanding of the effects of content moderation and clarify if and how automated systems deployed in a large online social network impact user behavior.
Further, our work can help clarify whether rule-breaking behavior is contagious.
Past work has typically relied on lab-based experimental settings to study the effect of rule-breaking or toxic content on subsequent comments~\cite{cheng2017anyone,kim2021distorting,han2015playing,rosner2016dangerous}. 
In contrast, here we examine the effect of \textit{removing} or \textit{hiding} rule-breaking content affects other users on a real social media platform, a setup with greater ecological validity.

This work is not the first in using quasi-experimental methods to evaluate the impact of moderation interventions (e.g., \cite{seering2017shaping,srinivasan2019content, teblunthuis2021effects}). However, approaches used in other work are not readily applicable to our scenario, \eg, they assume deterministic interventions~\cite{teblunthuis2021effects} or a random interval until an intervention is applied~\cite{srinivasan2019content}.
Thus, we propose a fuzzy regression discontinuity approach to estimate the effect of automated moderation on user behavior. 
This approach could be easily adapted to assess how other automated moderation systems affect subsequent user behavior.

\section{Background}
\label{sec:rwb}

\xhdr{Violence and incitement policy}
In this paper, we study a classifier and associated interventions used to help in enforcing Facebook's community standards for violence and incitement. The policy%
\footnote{\url{https://transparency.fb.com/policies/community-standards/violence-incitement/}. }
has the following rationale: \textit{``We aim to prevent potential offline harm that may be related to content on Facebook. While we understand that people commonly express disdain or disagreement by threatening or calling for violence in non-serious ways, we remove language that incites or facilitates serious violence. (\dots)''}

\xhdr{Interventions} In this paper, we study two interventions applied to rule-breaking content in the context of enforcing community guidelines.
These interventions, illustrated in  \Figref{fig:lead}, are applied incrementally.
Content whose score is greater than the first threshold $t_1$ is hidden.
Then, if the score crosses the second threshold $t_2$, it is immediately deleted, and a warning is sent to the offending user.
To other users, there is no indication that a post was made and later deleted. This approach aims to incrementally intervene upon content, acknowledging that some content that is borderline to community standards may remain in the social network with reduced visibility.%
\footnote{ \url{https://transparency.fb.com/features/approach-to-ranking/types-of-content-we-demote}}

\xhdr{Scope} The violence and incitement classifier studied here is only one of the ways that Facebook ensures that content follows community standards for violence and incitement. Other mechanisms also exist to ensure that content on Facebook adheres to these guidelines, and other community standards (e.g., for hate speech) are also enforced. These are beyond the scope of this paper.

\begin{table}
\caption{Outcomes considered in this study.}\label{tab:outcomes}

\begin{tabular}{@{}p{2.35cm}p{5cm}}
\toprule
\textbf{Outcome}           & \textbf{Description} \\ \midrule
Interventions in follow-up period & The number of  interventions that, in the follow-up period, targeted either the comments made by the user (in the user-level scenario) or the subsequent comments in the thread (in the thread-level scenario). \\ \midrule
Comments & The number of comments made during the follow-up period. In the user-level scenario, we also include posts. \\
\bottomrule
\end{tabular}
\end{table}

\section{Materials and Methods}
\label{sec:methods}

\begin{figure}[t]
    \centering

\begin{subfigure}[b]{\linewidth}
\centering

    \includegraphics[scale=0.9]{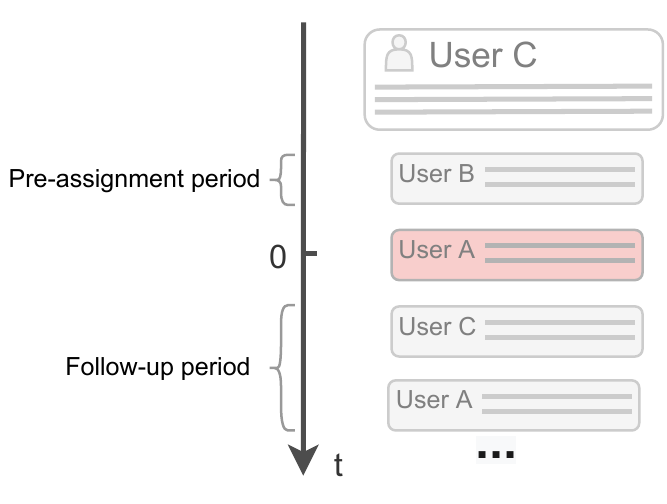}
    \caption{Thread-level scenario}
    \label{fig:exp1}
\end{subfigure}
\vspace{3mm}

\begin{subfigure}[b]{\linewidth}
\centering
    \includegraphics[scale=0.9]{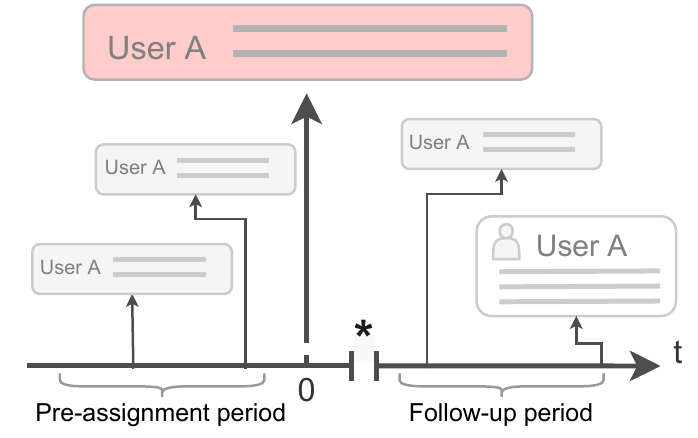}
    \caption{User-level scenario}
    \label{fig:exp2}
\end{subfigure}
\caption{The study approximates a real experiment where comments were intervened upon at random using observational data using a fuzzy regression discontinuity design. We depict the thread-level and user-level scenarios in (a) and (b) and describe them in  \Secref{sec:methods}. In (b), the asterisk denotes the setup where users are suspended, and the suspension period is not considered when calculating the outcomes.}
    \label{fig:exp}

\end{figure}

We studied the effect of automatically enforcing community guidelines with two quasi-experiments (\Figref{fig:exp}).
A \textit{post} is a piece of content posted on Facebook, a \textit{comment} is a response to that piece of content, and a \textit{thread} comprises comments associated with a post.

\xhdr{Data}
For both quasi-experiments, we used a dataset of public comments and posts posted by adult U.S. users in English between June 1st and August 31st, 2022. This comprised 412 million comments made in 1.5 million posts by 1.3 million distinct users.
All data was de-identified and analyzed in aggregate, and no individual-level data was viewed by the researchers.

\xhdr{Thread level} 
In this first scenario (\Figref{fig:exp1}), we studied the impact of automatic moderation on the thread where comments were intervened upon.
For each post in our data, we looked for the first comment $c_0$ whose score was in the 5 percentage point range of either of the two thresholds where the ``hide'' and the ``delete'' interventions are applied, \ie, $S \in [t_\text{hide} - 0.05, t_\text{hide} + 0.05]$ or $S \in [t_\text{delete} - 0.05, t_\text{delete} + 0.05]$; recall that $S \in [0,1]$.
(As described later, we reweight these data points based on their distance from $t_\text{hide}$ or $t_\text{delete}$.)
If a thread had no comments that met the above criteria, it was excluded from this analysis.
For each comment $c_0$ selected this way, we considered all comments made before (in the \textit{pre-assignment period}) and after $c_0$ in the same thread (in the \textit{follow-up period}).
After computing the outcome measures using data from the follow-up period, we used fuzzy regression discontinuity (see \Secref{sec:methods}) to determine the effect of hiding or deleting the comment.
To study effect heterogeneity, we considered four different setups in this quasi-experiment, varying (1) whether we included other comments from the author of the selected comment $c_0$ when calculating the outcomes of interest in the follow-up period; and (2) whether we considered threads that had more than $20$ comments. We choose 20 as a cutoff point as it induces an 80/20 split, \ie, around 80\% of the threads have less than 20 comments. 
A 75/25 or 85/15 split yielded qualitatively similar results.

\xhdr{User level} 
In the second scenario (\Figref{fig:exp2}), we studied the impact of automatic moderation on the users whose comments were intervened upon.
For each user $u$ in our data,  we looked for the first comment in the study period $c_0$ whose score was in the 5 percentage point range of the ``hide'' and ``delete'' thresholds.
If a user had no comments meeting the above criteria, they were excluded from this analysis.
For each user/comment tuple $(u_0,c_0)$ selected this way, we additionally considered all comments the user $u_0$ made in the $k$ days before (in the \textit{pre-assignment period}) and after posting $c_0$ (in the \textit{follow-up period}). 
Again, data from the follow-up period was used to calculate outcomes, and a fuzzy regression discontinuity design was used to determine the effect of the interventions.
We studied the heterogeneity of the effect of these interventions in two ways.
First, we varied the value of $k$, the number of days in the follow-up period (we considered $k \in \{7, 14, 21, 28\}$).
Second, we separately considered (1) users who had not violated community guidelines recently and only received a warning after having their comment deleted; and (2) users who had violated community guidelines once in the recent past and were thus suspended from posting on Facebook for a day after their comment was deleted.%
\footnote{\url{https://transparency.fb.com/en-gb/enforcement/taking-action/restricting-accounts/}}

\xhdr{Outcomes} The outcomes considered in this study are shown in \Tabref{tab:outcomes}. 
One outcome is associated with subsequent rule-breaking behavior (interventions), while one is associated with subsequent activity on the platform (comments).

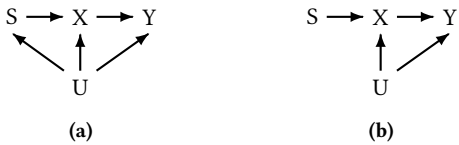
\begin{figure}[b]
\centering
\begin{subfigure}[b]{0.22\textwidth}
\centering
\begin{tikzpicture}
\node (1) {S};
\node [right = of 1] (2) {X};
\node [right = of 2] (3) {Y};
\node [below = of 2] (4) {U};

\draw[Arrow] (1.east) -- (2.west) ;
\draw[Arrow] (2.east) -- (3.west);
\draw[Arrow] (4.north west) -- (1.south);
\draw[Arrow] (4.north east) -- (3.south);
\draw[Arrow] (4.north) -- (2.south);
\end{tikzpicture} 
\subcaption{}
\label{fig:dag3}
\end{subfigure}
~
\begin{subfigure}[b]{0.22\textwidth}
\centering
\begin{tikzpicture}
\node (1) {S};
\node [right = of 1] (2) {X};
\node [right = of 2] (3) {Y};
\node [below = of 2] (4) {U};

\draw[Arrow] (1.east) -- (2.west) ;
\draw[Arrow] (2.east) -- (3.west);
\draw[Arrow] (4.north east) -- (3.south);
\draw[Arrow] (4.north) -- (2.south);
\end{tikzpicture} 
\subcaption{}
\label{fig:dag4}
\end{subfigure}
\caption{Causal Directed Acyclic Graphs (DAGs) illustrating the fuzzy regression discontinuity design. $S$ is the score attributed to a comment, $X$ is an indicator variable representing whether the comment was intervened upon, $U$ are unmeasured confounders, and $Y$ is the outcome of interest. 
While estimating the effect of $X$ on $Y$ is not possible in (a), around a specific threshold $t$ where there is a discontinuity around the probability of treatment ($P[X=1|S]$), we can remove the arrow $U \to X$ [see (b)] and use the same idea behind instrumental variable designs to measure the effect of $X$ on $Y$ (see main text).}
\label{fig:my_label}
\end{figure} 

\begin{figure}[t]
\centering
    \includegraphics[scale=.475]{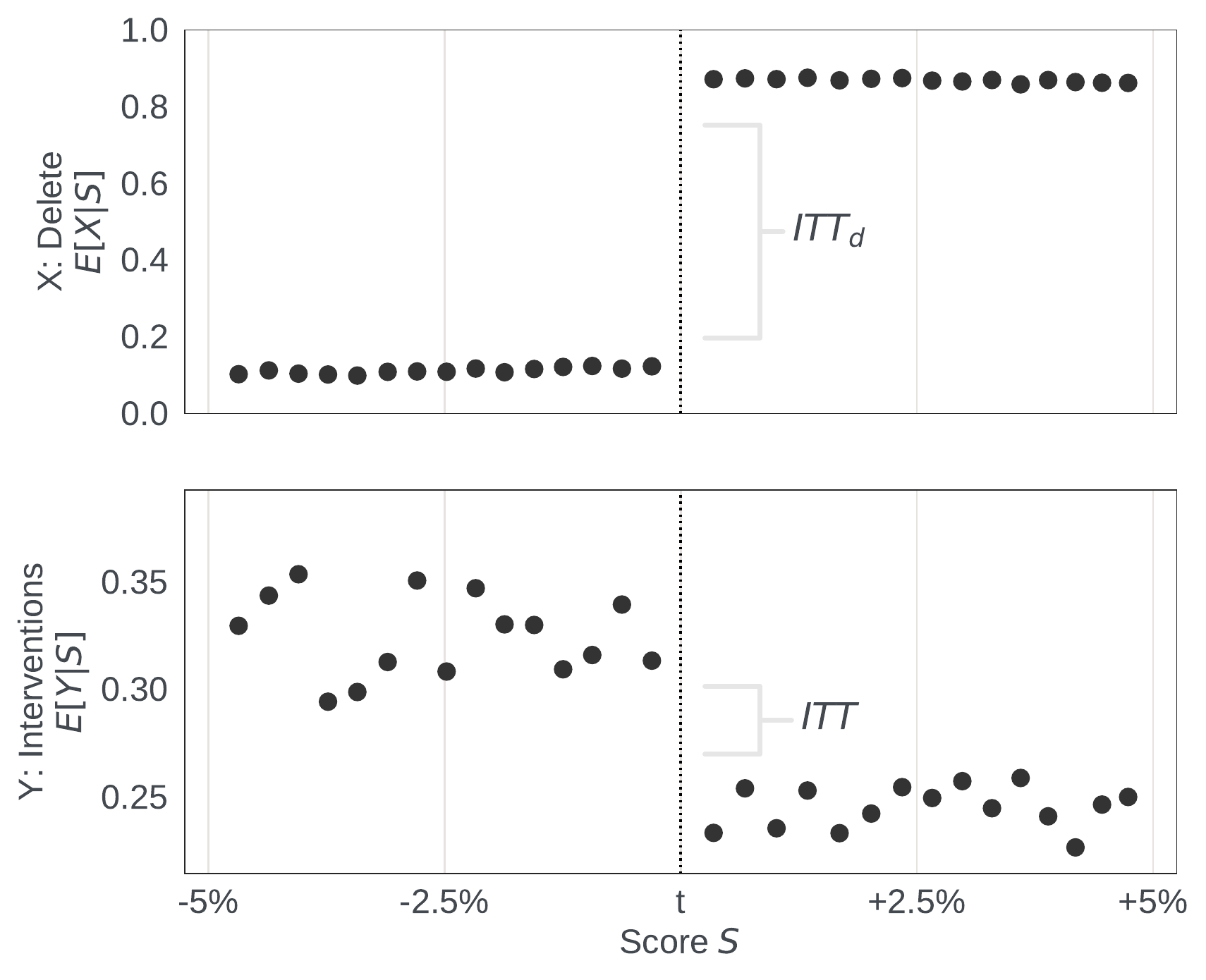}
\caption{A real example of our fuzzy regression discontinuity approach, considering the output of the violence and incitement classifier as the running variable $S$, deletions as the treatment $X$, and the number of interventions in a 7 day follow-up period as the outcome $Y$.
We estimate the causal effect as the ratio between two discontinuities ($ITT/ITT_d$).
$ITT_d$ (top figure) is the discontinuity in the treatment around the threshold $t$ (i.e., the probability of deletion), while $ITT$ (bottom figure) is the discontinuity in the outcome of interest around the same threshold (i.e., the number of interventions in a 7-day follow-up period).
}

\label{fig:expx}
\end{figure}

\begin{figure*}[t]
    \centering
    \includegraphics[width=\textwidth]{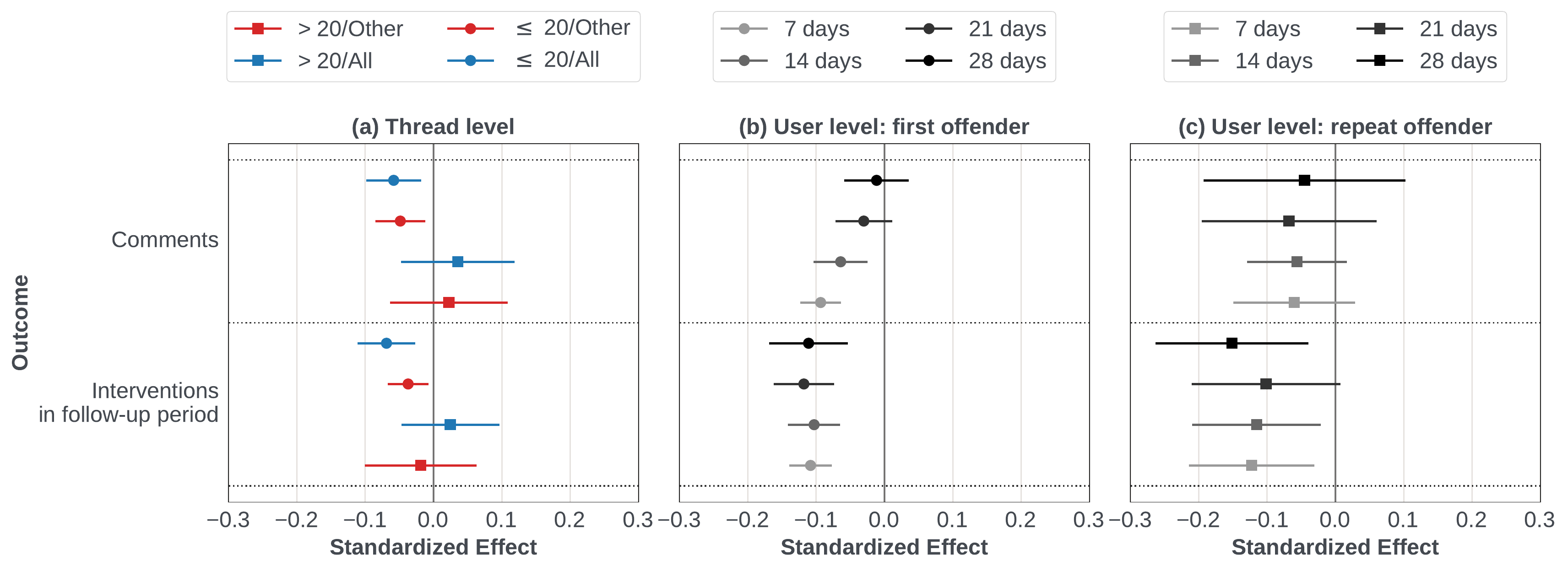}       
    \caption{We depict the estimated standardized effect of \emph{deleting} comments at the thread (a) and user level (b and c). Error bars represent 95\% CIs. We show that comment deletions can reduce subsequent activity (as measured by comments) and rule-breaking behavior (as measured by interventions in the follow-up period) across both (a) thread- and (b/c) user-level scenarios.}
    \label{fig:bigg}
\end{figure*}

\xhdr{Regression Discontinuity Designs}
Regression discontinuity (RD) is a  quasi-experimental study design that has been widely used in the social sciences since the 1990s~\cite{imbens2008regression}.
Here, we provide an overview of the fuzzy regression discontinuity design (an extension of RD), explaining how we use it for the quasi-experiments described in Section \ref{sec:methods} and illustrated in \Figref{fig:exp}.

\xhdr{Fuzzy Regression Discontinuity (FRD)} 
Let each comment $c$ be assigned a score $S_c \in [0,1]$, and $X_c$ be an indicator variable that equals 1 if the content has been intervened upon and 0 otherwise. 
If the $S$ score is beyond a threshold $t$, the probability of that comment getting intervened upon increases sharply:
\begin{equation}
    P[X_c = 1 \mid S_c] =
    \begin{cases}
    f_1(S_c) \text{ if } S_c \geq t \\
    f_0(S_c) \text{ if } S_c < t \\
    \end{cases}
    \text{where } f_1(a) > f_0(a) \quad \forall a.
\end{equation}
Note that this is a generalization of \textit{sharp} regression discontinuity designs, where $f_0(S_c) = 0$ and $f_1(S_c) = 1$ , \ie $P[X_c = 1 \mid S_c]$ jumps from 0 to 1 around the threshold. 
This is more suited to the scenario we are studying since mechanisms other than the classifier may come into play, \eg, comments may be removed due to user reports when the score is below the threshold ($S_c < t$), and other automated systems may prevent comments above the threshold from being removed when the score is above the threshold ($S_c \geq t$). 
A directed acyclic graph (DAG) illustrating the causal relationship between the score, the treatment, and the outcomes we are interested in measuring is shown in \Figref{fig:dag3}. The treatment $X$ is determined by the score $S$ given by the classifier and other factors unobserved in the present study (represented by $U$).

The key insight of fuzzy regression discontinuity designs is to estimate the effect of the intervention $X$ on the outcome $Y$, even with unknown confounders $U$, for comments with scores in the interval $S_c \in [t - \epsilon, t + \epsilon], \epsilon \to 0$.  
We assume that comments that lie right before or right after the threshold are indistinguishable, but those above the threshold are more likely to receive the treatment than those below.
Thus, around the threshold, we can consider a new DAG where there is no arrow $U \to S$, as shown in \Figref{fig:dag4}.
Here, $S$ has a causal effect on $Y$ only through $X$, and thus we can use the same idea behind instrumental variable (IV) designs \cite{angrist2008getting} to study the effect of $X$ on $Y$.
In IV designs, we estimate the Local Average Treatment Effect (LATE), the treatment effect for the subset of the comments that take the treatment (\ie, $X_c = 1)$ if and only if they were ``assigned'' to the treatment (\ie, $S_c > t$):
\begin{equation}
\label{eq2}
LATE = \frac{ITT}{ITT_d},
\end{equation}
where ITT is the average effect of assigning comments to the treatment group (regardless of them being treated), and $ITT_d$ is the proportion of subjects treated when assigned to the treated group.
As $S$ is only an instrument close to the threshold $t$, we estimate the LATE at the cutoff point (LATEC), rewriting Equation~\eqref{eq2} as:
\begin{equation}
\label{eq3}
LATEC = \frac{
{
E[Y_c \mid S_c = t + \epsilon]  - E[Y_c \mid S_c = t - \epsilon] }}
{
{
 E[X_c \mid S_c = t + \epsilon]  - E[X_c \mid S_c = t - \epsilon] }
  },~\epsilon \to 0.
\end{equation}

In practice, we can estimate the LATEC with 2-stage least squares regression, \ie, regressing the treatment $X$ on the score $S$ (first-stage), and then the outcome $Y$ on the values $\hat{X}$ predicted on the first-stage (second-stage), see \cite{angrist1995two} for details.
However, we do not have infinite data, and we cannot consider only comments with $S_c \in [t-\epsilon, t+\epsilon],~\epsilon \to 0$. This creates a bias--variance trade-off in the estimation of the LATEC. 
On the one hand, the wider the range we consider around the threshold $t$, the more the unmeasured confounders can bias our estimator.
On the other hand, the narrower the range, the less data we have, and thus the larger the variance of our estimator.

A common solution to navigating this trade-off consists of using a local linear regression~\cite{hahn2001identification},  where data points (here, comments) receive importance proportional to how far they are from the threshold, using a triangular weighting kernel defined as
\begin{equation}
\label{eq4}
K(S) = \mathbf{1}_{|S - t| < h} \Big(1 - \frac{S - t}{h}\Big),
\end{equation}
where $h$ is the bandwidth of the kernel that controls the bias--variance trade-off, and $\mathbf{1}_{|S - t| < h}$ is an indicator variable that equals 1 if $|S - t| < h$ and 0 otherwise. We empirically determine the bandwidth $h$, choosing the bandwidth that yields the optimal mean squared error (MSE) of the LATEC estimator~\cite{imbens2012optimal}.

\xhdr{Example} \Figref{fig:expx} 
illustrates our fuzzy regression discontinuity design. It uses a random sample of users who did not previously violate community guidelines and examines interventions in a 7-day follow-up period following the first comment of interest ($c_0$).
Figure~\ref{fig:expx} (top) shows the percentage of first comments ($c_0$) that received the ``Delete'' treatment (\ie $E[X | S]$; in the $y$-axis) for different scores received by first comments $c_0$ (in the $x$-axis).
Figure~\ref{fig:expx} (bottom) depicts the outcome ``Interventions in the follow-up period'' ($E[Y | S]$; in the $y$-axis) for different scores by first comments $c_0$  (in the $x$-axis).
Intuitively, the regression discontinuity design estimates the treatment effect of $X$ on $Y$ around the threshold $t_\text{delete}$ by dividing the discontinuity in $E[Y | S]$ [corresponding to the numerator in Eq.~\eqref{eq2} and  Eq.~\eqref{eq3}]   by the discontinuity in
$E[X | S]$ [corresponding to the denominator in Eq.~\eqref{eq2} and  Eq.~\eqref{eq3}]. 

\xhdr{Robustness checks} To ensure the validity of our regression discontinuity design, we additionally conduct several robustness checks suggested by guides outlining best practices~\cite{lee2010regression,jacob2012practical}. These robustness checks can be found in Appendix~\ref{app:rc}.

\section{Results}

Using FRD, we estimated the effect of the ``Hide'' and ``Delete'' interventions for the thread- and user-level scenarios. We depict the standardized effects associated with our key findings in \Figref{fig:bigg} and present all the estimated effects in \Tabref{tab:all}.

\subsection{Thread level}
\Figref{fig:bigg}(a) shows the standardized effect of deleting comments on the number of comments and interventions in the follow-up period in the thread-level scenario).
Comment deletion had a significant effect on both the number of subsequent interventions and the number of subsequent comments in threads that had fewer than 20 posts prior to the intervention.
When comments from the original commenter were included ($\mathbf{\leq}$\textbf{20/All}), the intervention reduced the number of comments by $-13.16$ (95\% CI: $-21.23$, $-5.10$) and the number of subsequent interventions by $-0.946$ (95\% CI: $-1.59$, $-0.299$; see non-standardized effects shown in \Tabref{tab:all}).
To get a sense of the effect size, we calculated the average number of comments and interventions in the follow-up period received by threads right below the intervention threshold, where $S \in [t_\text{delete} - 0.01, t_\text{delete})$.
Threads in the $\mathbf{\leq}$\textbf{20/All} scenario just before the threshold received $1.5$ interventions (95\% CI: $1.3$, $1.9$) and $27.7$ comments (95\% CI: $23.2$, $33.,2$) on average, suggesting that these effects were substantial.

Still considering the \textbf{All} scenario, for both outcomes, the effect of deletions was neither substantial nor significant for threads that already had more than 20 comments when the delete intervention happened [\eg, see $\mathbf{>}$\textbf{20/All} in \Figref{fig:bigg}(a)].
Deleting any single comment may have less of an effect in longer threads because participants are less likely to see such a comment (e.g., because such a comment may already have been hidden or because there are at least 19 other comments to see).
Effects may also have been more difficult to observe because of the smaller sample size---there were fewer threads with more than 20 comments than threads with 20 or fewer comments.
Hiding comments, as opposed to deleting, had small and statistically insignificant effects on the number of subsequent interventions and  comments for all setups considered (see \Tabref{tab:all}; we discuss this further in \Secref{sec:disc}). 

We additionally computed the same outcomes in the follow-up period, not considering comments by the original commenter to understand if the effect was due to changes in the behavior of the original commenter (\ie, the individual who had their comment intervened upon) or other users in the thread (scenarios $\mathbf{\leq}$\textbf{20/Other} and $\mathbf{>}$\textbf{20/Other} in \Figref{fig:bigg}(a) and \Tabref{tab:all}).
We found that standardized effects remained qualitatively similar, \eg comments were reduced $-0.069$ standard deviations (SDs) in the $\mathbf{\leq}$\textbf{20/All} setup \vs  $-0.037$ SDs in the $\mathbf{\leq}$\textbf{20/Other} setup, suggesting that the intervention discouraged \textit{other users} from posting rule-breaking comments.

\subsection{User level}
For the user-level scenario, we considered both cases where users did and did not have comments deleted previously. 
We make this distinction as the interventions for these users differ: ``first-time offenders'' only receive a warning, whereas ``repeat offenders'' additionally have their posting privileges suspended for 24 hours. 
The suspension period for repeat offenders is not considered in  the follow-up period as it could explain behavior differences.

\xhdr{User level: first-time offenders}
\Figref{fig:bigg}(b) shows the effect of deletions for first-time offenders. 
Again, deleting comments had significant effects on both outcomes.
Considering the 7 days following the intervention, deletion decreased the number of comments by $4.6$ and decreased the number of subsequent interventions by $0.12$.
To get a sense of the effect size, we calculated outcomes for users right below the intervention threshold, $S \in [t_\text{delete} - 0.01, t_\text{delete})$.
These users received on average  $0.23$ interventions in the follow-up period (95\% CI: $0.21$, $0.25$) and made on average $13$ comments (95\% CI: $12.6$, $13.4$), suggesting that the effects are substantial.
Setups that considered larger intervention periods (21 and 28 days) showed that, while the effect on the subsequent number of comments waned with time (\ie, effects were smaller for longer follow-up periods), the effect on the number of subsequent interventions was largely stable.
This indicates that automated content moderation has positive, long-lasting effects on subsequent rule-breaking behavior.
Hiding comments had small and statistically insignificant effects on the number of subsequent interventions and comments. 

\xhdr{User level: repeat offenders}
\Figref{fig:bigg}(c) shows the effect of deletions for repeat offenders. 
For these users, deleting comments yielded decreases in both the number of interventions and comments in the follow-up period. 
The wider confidence intervals here may be partially explained by the smaller sample, as fewer users had their comments deleted a second time.
Nonetheless, for 3 out of the 4 time periods considered (7, 14, 28 days), we again observed significant effects that were similar in magnitude to the effects observed in the ``first-time offender'' setup.
Considering a 28-day follow-up period, deletions decreased interventions received by repeat offenders by $0.28$ (95\% CI: $-0.48$, $-0.078$) \vs $0.192$ for first-time offenders.
This suggests that deletions are also effective for users who have previously broken community guidelines.
Hiding comments had small and statistically insignificant effects on the considered outcomes.

\section{Discussion and conclusion}
\label{sec:disc}

Content moderation systems are essential to the functioning of mainstream social networks~\cite{gillespie2018custodians} and can prevent harm by removing rule-breaking content before anyone sees or interacts with it~\cite{grimmelmann2015virtues}.
In this work, we studied how these systems may also positively impact on-platform user behavior.
Using a fuzzy regression discontinuity design~\cite{imbens2008regression}, we found that comment deletion had substantial and statistically significant effects on subsequent rule-breaking behavior and user activity. 
At the user level, for ``first-time offenders,'' deletions had long-lasting effects on reducing rule breaking, but only temporary effects on posting activity, suggesting that comment moderation does not necessarily require making a trade-off between safety and engagement.
This result is qualitatively aligned with the findings of Srinivasan et al.~\cite{srinivasan2019content} on the r/ChangeMyView community on Reddit and suggests that automated platform-level moderation may yield the same effects as manual community-level moderation.
At the thread level, we found that content moderation reduced rule-breaking activity even for other users who were not intervened upon.
This result is qualitatively aligned with previous work suggesting that uncivil behavior is contagious~\cite{cheng2017anyone,kim2021distorting}, further highlighting the importance of proactive content moderation.

We also found that hiding comments did not have substantial or significant effects.
This may be linked to an important limitation of our work: we were able to measure the effect of content moderation only at the thresholds at which they were applied. The hiding intervention may have a stronger effect at a different threshold.
Importantly, this study does not necessarily imply that comment hiding is not useful, as hiding comments can still prevent harm by reducing exposure to borderline content and may have other beneficial effects that we did not measure.
In that context, future work could also find ways to estimate the effect of moderation across various thresholds.
At the same time, the effects of deletion reported here may also be an underestimate.
As interventions on Facebook are ``cumulative,'' when we study the effect of deletion, we do not compare ``deletion'' with ``no deletion,'' but instead can only compare ``deletion'' with ``hiding.''
Therefore, it could be that the effect of deleting content is even stronger, but that part of the effect is masked by the ``hiding'' intervention (which, as previously stated, might itself be impactful if enacted at higher thresholds).
Last, our study is also limited in that we consider specific interventions enacted only upon U.S.-based Facebook users, with effects that could be heterogeneous across other platforms and countries. 
Despite the aforementioned limitations, we argue that, even in this specific setting, understanding the impact of in-production content moderation systems is of great importance as a first step toward a more holistic understanding of how automated moderation systems impact online platforms such as Facebook.

Last, we argue that the methodology discussed and applied in this paper can be used to assess moderation interventions across different scenarios and platforms.
While much of the literature on harmful content has focused on developing methods to accurately detect such content, here we provide a way to measure the effects of deploying these systems (and their associated interventions) on our information ecosystem.

\xhdr{Acknowledgements} 
We thank the Core Data Science team at Meta for the useful discussions.
West's lab is partly supported by grants from
Swiss National Science Foundation (200021\_185043),
Swiss Data Science Center (P22\_08),
H2020 (952215),
Microsoft Swiss Joint Research Center,
and Google,
and by generous gifts from Facebook, Google, and Microsoft.

\begin{table*}
    \centering
    \caption{Summary of the effects across all settings. For the thread-level scenario, the setups where we consider only threads with less than 20 comments are marked with $\leq$ 20 in the ``Setup'' column (\vs $>$20 for the setup considering more than 20 comments), and the setups where the original commenter is not considered are marked as ``Other'' (\vs ``All'' for when they are). For the user-level scenario, the ``Setup'' column shows the number of days considered in the follow-up period. Stars $*$ indicate statistically significant effects, \ie $p<0.05$}\label{tab:all}
\begin{tabular}{lll|lllr}
\toprule
     &                              &               &    &                       Effect &        Effect (Standardized) &       n \\
Intervention & Scenario & Outcome & Setup &                              &                              &         \\
\midrule
Delete & Thread-level & Comments & $\leq$20 / All &   -13.16 (-21.23, -5.10)$^*$ &  -0.069 (-0.111, -0.027)$^*$ &  190885 \\
     &                              &               & $\leq$20 / Other &    -7.41 (-13.39, -1.42)$^*$ &  -0.037 (-0.067, -0.007)$^*$ &  200655 \\
     &                              &               & >20 / All &       43.22 (-82.98, 169.41) &        0.025 (-0.047, 0.096) &   49645 \\
     &                              &               & >20 / Other &     -34.03 (-183.04, 114.97) &       -0.019 (-0.101, 0.063) &   52241 \\
     &                              & Interventions & $\leq$20 / All &   -0.946 (-1.59, -0.299)$^*$ &  -0.058 (-0.098, -0.018)$^*$ &  190885 \\
     &                              &               & $\leq$20 / Other &   -0.876 (-1.53, -0.218)$^*$ &  -0.049 (-0.085, -0.012)$^*$ &  200655 \\
     &                              &               & >20 / All &           2.27 (-3.02, 7.56) &        0.036 (-0.048, 0.119) &   49645 \\
     &                              &               & >20 / Other &           1.39 (-3.97, 6.74) &        0.022 (-0.064, 0.109) &   52241 \\
     & User-level (first offender) & Comments & 7 &     -4.55 (-6.00, -3.11)$^*$ &  -0.093 (-0.123, -0.064)$^*$ &  162149 \\
     &                              &               & 14 &     -5.72 (-9.25, -2.19)$^*$ &  -0.064 (-0.104, -0.025)$^*$ &  112793 \\
     &                              &               & 21 &          -3.95 (-9.38, 1.48) &       -0.030 (-0.072, 0.011) &   84592 \\
     &                              &               & 28 &         -1.99 (-10.12, 6.14) &       -0.012 (-0.059, 0.036) &   60175 \\
     &                              & Interventions & 7 &  -0.117 (-0.151, -0.084)$^*$ &  -0.108 (-0.139, -0.077)$^*$ &  162149 \\
     &                              &               & 14 &  -0.144 (-0.197, -0.090)$^*$ &  -0.103 (-0.141, -0.065)$^*$ &  112793 \\
     &                              &               & 21 &  -0.196 (-0.270, -0.123)$^*$ &  -0.118 (-0.162, -0.074)$^*$ &   84592 \\
     &                              &               & 28 &  -0.192 (-0.291, -0.092)$^*$ &  -0.111 (-0.169, -0.053)$^*$ &   60175 \\
     & User-level (repeat offender) & Comments & 7 &          -3.37 (-8.35, 1.61) &       -0.060 (-0.149, 0.029) &   29825 \\
     &                              &               & 14 &         -6.11 (-14.02, 1.81) &       -0.056 (-0.129, 0.017) &   26596 \\
     &                              &               & 21 &        -11.07 (-31.95, 9.81) &       -0.068 (-0.196, 0.060) &   21693 \\
     &                              &               & 28 &        -9.84 (-42.03, 22.34) &       -0.045 (-0.193, 0.103) &   18468 \\
     &                              & Interventions & 7 &  -0.107 (-0.187, -0.027)$^*$ &  -0.122 (-0.214, -0.031)$^*$ &   29825 \\
     &                              &               & 14 &  -0.155 (-0.281, -0.029)$^*$ &  -0.115 (-0.209, -0.021)$^*$ &   26596 \\
     &                              &               & 21 &       -0.166 (-0.344, 0.012) &       -0.101 (-0.210, 0.007) &   21693 \\
     &                              &               & 28 &  -0.277 (-0.483, -0.072)$^*$ &  -0.151 (-0.263, -0.039)$^*$ &   18468 \\ \midrule
Hide & Thread-level & Comments & $\leq$20 / All &          0.718 (-1.93, 3.36) &        0.004 (-0.011, 0.019) &  868632 \\
     &                              &               & $\leq$20 / Other &          0.928 (-1.31, 3.17) &        0.005 (-0.007, 0.018) &  907871 \\
     &                              &               & >20 / All &        -8.90 (-67.40, 49.60) &       -0.003 (-0.026, 0.019) &  300716 \\
     &                              &               & >20 / Other &       -10.94 (-68.34, 46.45) &       -0.004 (-0.025, 0.017) &  314287 \\
     &                              & Interventions & $\leq$20 / All &        0.023 (-0.096, 0.142) &        0.003 (-0.011, 0.017) &  868632 \\
     &                              &               & $\leq$20 / Other &        0.049 (-0.065, 0.163) &        0.006 (-0.008, 0.019) &  907871 \\
     &                              &               & >20 / All &         0.285 (-0.492, 1.06) &        0.011 (-0.018, 0.040) &  300716 \\
     &                              &               & >20 / Other &        0.241 (-0.492, 0.974) &        0.009 (-0.018, 0.036) &  314287 \\
     & User-level (first offender) & Comments & 7 &        -0.568 (-1.41, 0.272) &       -0.010 (-0.024, 0.005) &  723278 \\
     &                              &               & 14 &         -1.19 (-2.90, 0.526) &       -0.011 (-0.027, 0.005) &  542780 \\
     &                              &               & 21 &          -1.98 (-4.97, 1.00) &       -0.013 (-0.031, 0.006) &  422996 \\
     &                              &               & 28 &         -0.350 (-6.11, 5.41) &       -0.002 (-0.029, 0.026) &  291712 \\
     &                              & Interventions & 7 &        0.007 (-0.011, 0.026) &        0.008 (-0.012, 0.027) &  723278 \\
     &                              &               & 14 &        0.005 (-0.015, 0.025) &        0.004 (-0.012, 0.020) &  542780 \\
     &                              &               & 21 &        0.018 (-0.015, 0.050) &        0.012 (-0.010, 0.033) &  422996 \\
     &                              &               & 28 &        0.032 (-0.007, 0.071) &        0.019 (-0.004, 0.041) &  291712 \\
     & User-level (repeat offender) & Comments & 7 &          0.948 (-1.11, 3.01) &        0.014 (-0.017, 0.045) &  126587 \\
     &                              &               & 14 &           2.68 (-1.44, 6.79) &        0.021 (-0.011, 0.053) &  122545 \\
     &                              &               & 21 &          3.30 (-3.64, 10.24) &        0.017 (-0.019, 0.053) &  103467 \\
     &                              &               & 28 &          3.92 (-6.41, 14.24) &        0.015 (-0.025, 0.056) &   82067 \\
     &                              & Interventions & 7 &       -0.002 (-0.034, 0.029) &       -0.003 (-0.039, 0.033) &  126587 \\
     &                              &               & 14 &        0.017 (-0.025, 0.059) &        0.013 (-0.018, 0.044) &  122545 \\
     &                              &               & 21 &        0.013 (-0.066, 0.092) &        0.008 (-0.039, 0.054) &  103467 \\
     &                              &               & 28 &        0.000 (-0.113, 0.113) &        0.000 (-0.059, 0.060) &   82067 \\
\bottomrule
\end{tabular}
\end{table*}

\bibliographystyle{ACM-Reference-Format}
\bibliography{sample-base}

\appendix

\begin{figure}
    \centering
    \includegraphics[width=\linewidth]{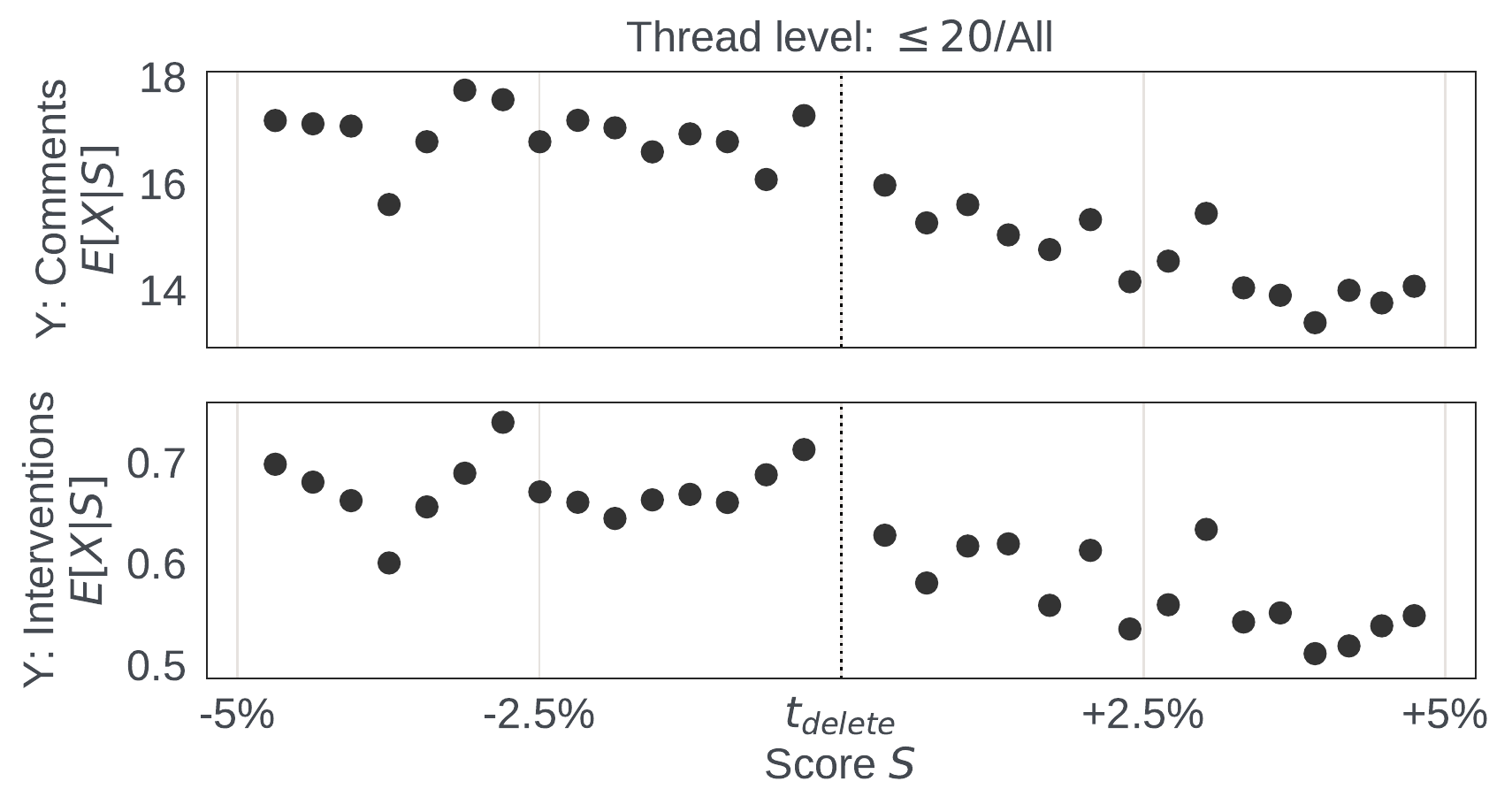}
    \caption{Example of the discontinuities in the outcome (top: comments; bottom: interventions) we visually inspected to ensure the validity of our approach.}
    \label{figs:expx2}
\end{figure}
    
\begin{figure}
    \centering
    \includegraphics[width=\linewidth]{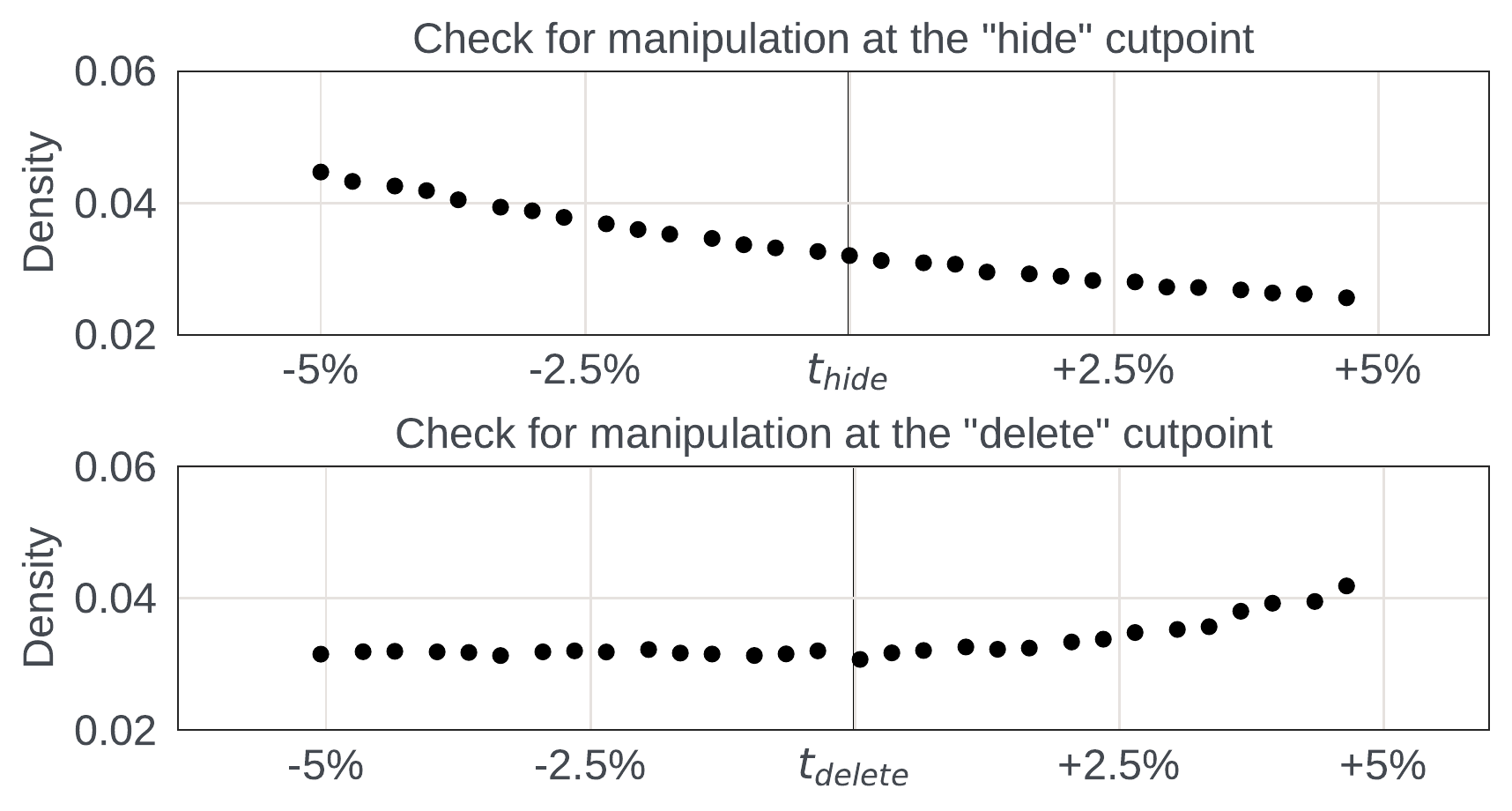}
    \caption{Density of the running variable (\ie, the score $S$) around the thresholds where content gets deleted ($t_\text{delete}$) and hidden ($t_\text{hide}$).}
    \label{figs:manip}
\end{figure}

\begin{figure}[t]
    \includegraphics[width=\linewidth]{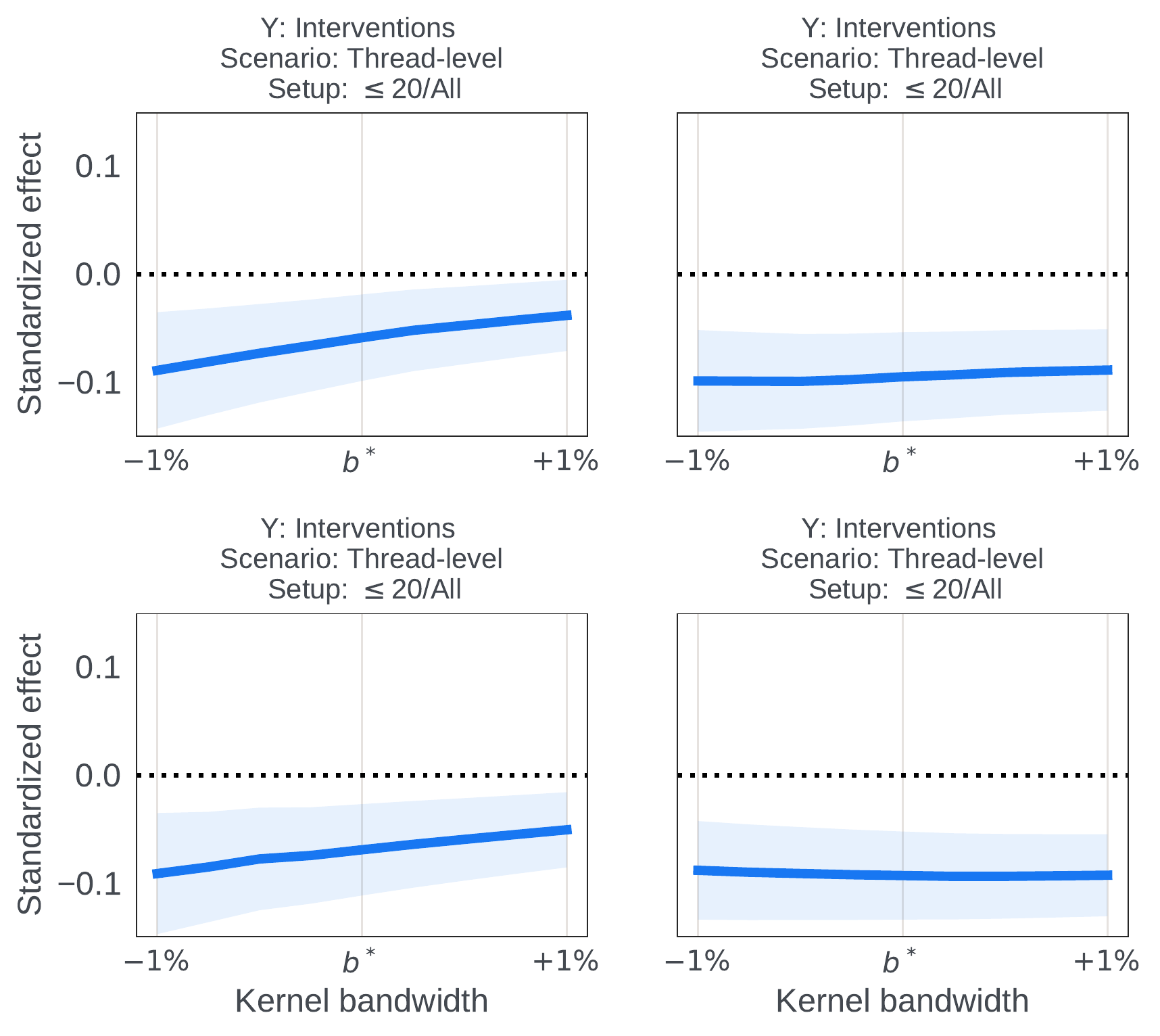}
    \caption{We show how the standardized effect ($y$-axis) of four of the regression discontinuity designs vary with slight changes to the kernel bandwidth. $b^*$ corresponds to the MSE-optimal bandwidth for each FRD (as described in ~\cite{imbens2012optimal}).}
    \label{figs:th}
\end{figure}

\section{Robustness Checks}\label{app:rc}

\xhdr{Visual analysis} As a first sanity check, we visually inspect discontinuities in the outcome variables around the thresholds (\eg, as shown in \Figref{fig:expx}). 
We find that we can visualize the discontinuities right around the threshold, as expected in a fuzzy regression discontinuity design.
In \Figref{figs:expx2}, we further show the discontinuities for the thread-level scenario (setup: $\mathbf{\leq}$\textbf{20/All}).

\xhdr{Manipulation at the cutpoint} One well-established threat to the validity of RD designs is that individuals may have knowledge about the cutpoint and adjust the running variable (for us, the score $S$) to fall right before or right above it. For instance, in our scenario, if users knew exactly what score their comment would receive before posting it, they might re-word until they find it below the threshold. While in our case, we do not consider this threat to be credible, we entertain the hypothesis and conduct the standard robustness checks, inspecting the density of scores around the threshold and conducting the McCrary test  \cite{mccrary2008manipulation}  to assess whether the discontinuity in the density of the rating variable at the cutpoint equals zero. We plot the density in \Figref{figs:manip}, which shows no indication of manipulation around the threshold (as does the McCrary test, where $p>0.05$).

\xhdr{Placebo FRDs} A key assumption of FRD is that around the threshold, units are exactly the same except for the fact that those above the threshold have an increased chance of receiving the treatment. 
As such, it is commonplace (\eg, see \cite{jacob2012practical}) to repeat the entire FRD analysis considering a variable that the treatment should not impact. 
If comments below and above the threshold are comparable, we should not see significant differences for these placebo FRDs. 
In our case, we run placebo FRDs considering the same outcome variables calculated in the pre-assignment period (see \Figref{fig:exp}), where no intervention occurred. Thus, we should expect no discontinuity in the outcomes. Results are reported in \Tabref{tab:placebo}. We find that effects in the placebo FRDs are small and not statistically significant, \ie, $p > 0.05$, suggesting that, indeed, comments below and above the threshold are comparable.

\xhdr{Varying the bandwidth} Finally, to ensure our findings were robust to slight changes in the kernel bandwidth, we repeated the FRDs with varying bandwidth sizes. In \Figref{figs:th} we show the changes in the standardized effect for four of the FRDs carried (at both the user level and the thread level) when varying the kernel bandwidth around the MSE-optimal bandwidth $b^*$. 
Overall, we find that our results are robust to slight changes in the kernel bandwidth.

\begin{table*}
\caption{Results for placebo FRDs. This table is exactly the same as \Tabref{tab:all} except that outcomes are calculated in the pre-assignment period and thus should not differ between treatment and control groups (\ie, those whose comment fell above or below the intervention thresholds).}
\label{tab:placebo}
    \centering
\begin{tabular}{lll|lllr}
\toprule
     &                              &               &    &                  Effect &   Effect (Standardized) &       n \\
Intervention & Scenario & Outcome & Setup &                         &                         &         \\
\midrule
Delete & Thread-level & Comments & $\leq$20 / All &  -0.125 (-0.319, 0.068) &  -0.024 (-0.062, 0.013) &  190885 \\
     &                              &               & $\leq$20 / Other &  -0.153 (-0.336, 0.031) &  -0.030 (-0.065, 0.006) &  200655 \\
     &                              &               & >20 / All &  -11.87 (-41.36, 17.62) &  -0.022 (-0.076, 0.033) &   49645 \\
     &                              &               & >20 / Other &   -20.34 (-49.02, 8.33) &  -0.046 (-0.111, 0.019) &   52241 \\
     &                              & Interventions & $\leq$20 / All &   0.003 (-0.011, 0.016) &   0.006 (-0.022, 0.034) &  190885 \\
     &                              &               & $\leq$20 / Other &   0.001 (-0.013, 0.015) &   0.002 (-0.028, 0.032) &  200655 \\
     &                              &               & >20 / All &   0.038 (-0.429, 0.505) &   0.005 (-0.061, 0.072) &   49645 \\
     &                              &               & >20 / Other &  -0.005 (-0.450, 0.441) &  -0.001 (-0.067, 0.065) &   52241 \\
     & User-level (first offender) & Comments & 7 &    -0.186 (-1.77, 1.39) &  -0.004 (-0.038, 0.030) &  162149 \\
     &                              &               & 14 &    -0.545 (-3.38, 2.29) &  -0.007 (-0.045, 0.031) &  112793 \\
     &                              &               & 21 &     0.569 (-4.11, 5.25) &   0.006 (-0.042, 0.054) &   84592 \\
     &                              &               & 28 &      3.68 (-2.35, 9.71) &   0.029 (-0.019, 0.078) &   60175 \\
     &                              & Interventions & 7 &   0.013 (-0.008, 0.035) &   0.019 (-0.012, 0.050) &  162149 \\
     &                              &               & 14 &   0.001 (-0.029, 0.031) &   0.001 (-0.036, 0.039) &  112793 \\
     &                              &               & 21 &  -0.012 (-0.052, 0.029) &  -0.013 (-0.057, 0.031) &   84592 \\
     &                              &               & 28 &  -0.008 (-0.065, 0.049) &  -0.008 (-0.063, 0.048) &   60175 \\
     & User-level (repeat offender) & Comments & 7 &    -0.369 (-5.73, 4.99) &  -0.007 (-0.108, 0.094) &   29825 \\
     &                              &               & 14 &    -0.906 (-7.57, 5.76) &  -0.010 (-0.081, 0.061) &   26596 \\
     &                              &               & 21 &    1.94 (-12.45, 16.33) &   0.015 (-0.095, 0.124) &   21693 \\
     &                              &               & 28 &    2.01 (-15.87, 19.90) &   0.012 (-0.097, 0.121) &   18468 \\
     &                              & Interventions & 7 &   0.048 (-0.022, 0.118) &   0.057 (-0.026, 0.139) &   29825 \\
     &                              &               & 14 &   0.036 (-0.069, 0.141) &   0.031 (-0.061, 0.124) &   26596 \\
     &                              &               & 21 &   0.086 (-0.047, 0.219) &   0.062 (-0.034, 0.159) &   21693 \\
     &                              &               & 28 &   0.021 (-0.128, 0.170) &   0.013 (-0.081, 0.108) &   18468 \\ \midrule
Hide & Thread-level & Comments & $\leq$20 / All &  -0.054 (-0.150, 0.042) &  -0.010 (-0.029, 0.008) &  868632 \\
     &                              &               & $\leq$20 / Other &  -0.039 (-0.140, 0.062) &  -0.007 (-0.027, 0.012) &  907871 \\
     &                              &               & >20 / All &   -18.16 (-40.09, 3.77) &  -0.018 (-0.041, 0.004) &  300716 \\
     &                              &               & >20 / Other &   -17.14 (-38.37, 4.08) &  -0.017 (-0.037, 0.004) &  314287 \\
     &                              & Interventions & $\leq$20 / All &   0.001 (-0.005, 0.006) &   0.002 (-0.014, 0.019) &  868632 \\
     &                              &               & $\leq$20 / Other &   0.001 (-0.005, 0.006) &   0.002 (-0.014, 0.018) &  907871 \\
     &                              &               & >20 / All &  -0.021 (-0.146, 0.104) &  -0.005 (-0.033, 0.023) &  300716 \\
     &                              &               & >20 / Other &  -0.063 (-0.163, 0.038) &  -0.014 (-0.036, 0.008) &  314287 \\
     & User-level (first offender) & Comments & 7 &   -0.501 (-1.28, 0.279) &  -0.009 (-0.023, 0.005) &  723278 \\
     &                              &               & 14 &   -0.972 (-2.37, 0.422) &  -0.011 (-0.027, 0.005) &  542780 \\
     &                              &               & 21 &    -0.877 (-3.07, 1.32) &  -0.008 (-0.026, 0.011) &  422996 \\
     &                              &               & 28 &      1.30 (-2.53, 5.14) &   0.008 (-0.016, 0.033) &  291712 \\
     &                              & Interventions & 7 &   0.003 (-0.008, 0.015) &   0.006 (-0.014, 0.025) &  723278 \\
     &                              &               & 14 &  -0.009 (-0.025, 0.008) &  -0.013 (-0.037, 0.011) &  542780 \\
     &                              &               & 21 &  -0.011 (-0.033, 0.010) &  -0.014 (-0.042, 0.013) &  422996 \\
     &                              &               & 28 &  -0.003 (-0.024, 0.018) &  -0.003 (-0.026, 0.020) &  291712 \\
     & User-level (repeat offender) & Comments & 7 &     0.231 (-2.11, 2.58) &   0.004 (-0.033, 0.040) &  126587 \\
     &                              &               & 14 &     0.766 (-3.10, 4.63) &   0.007 (-0.028, 0.042) &  122545 \\
     &                              &               & 21 &     0.214 (-5.26, 5.69) &   0.001 (-0.034, 0.037) &  103467 \\
     &                              &               & 28 &     3.83 (-3.82, 11.49) &   0.020 (-0.020, 0.060) &   82067 \\
     &                              & Interventions & 7 &   0.003 (-0.021, 0.026) &   0.004 (-0.028, 0.036) &  126587 \\
     &                              &               & 14 &   0.002 (-0.035, 0.038) &   0.002 (-0.035, 0.038) &  122545 \\
     &                              &               & 21 &   0.029 (-0.038, 0.096) &   0.024 (-0.032, 0.079) &  103467 \\
     &                              &               & 28 &   0.009 (-0.047, 0.065) &   0.007 (-0.034, 0.047) &   82067 \\
\bottomrule
\end{tabular}
\end{table*}

\end{document}

\endinput

%% file: dlabmacros.tex
\usepackage{hyphenat}
\usepackage{xspace}

\usepackage{amsthm}
\theoremstyle{plain}

\newcommand{\abbrevStyle}[1]{#1}

\newcommand{\ie}{\abbrevStyle{i.e.}\xspace}
\newcommand{\eg}{\abbrevStyle{e.g.}\xspace}

\newcommand{\vs}{\abbrevStyle{vs.}\xspace}

\newcommand{\Secref}[1]{Sec.~\ref{#1}}

\newcommand{\Tabref}[1]{Table~\ref{#1}}
\newcommand{\Figref}[1]{Fig.~\ref{#1}}

\newcommand{\xhdr}[1]{\vspace{0.7mm}\noindent{{\bf #1.}}}

\newcommand{\textcite}[1]{\citeauthor{#1} \shortcite{#1}}

\newcommand{\hide}[1]{}

\makeatletter
\newcommand{\iffont}[2]{\ifthenelse{\equal{\f@family}{#1}}{#2}{}}
\makeatother

\iffont{ptm}{
  \usepackage{mathptmx}

  \DeclareSymbolFont{greek}{OML}{cmm}{m}{n}
  \DeclareMathSymbol{\alpha}{\mathalpha}{greek}{"0B}
  \DeclareMathSymbol{\beta}{\mathalpha}{greek}{"0C}
  \DeclareMathSymbol{\gamma}{\mathalpha}{greek}{"0D}
  \DeclareMathSymbol{\delta}{\mathalpha}{greek}{"0E}
  \DeclareMathSymbol{\epsilon}{\mathalpha}{greek}{"0F}
  \DeclareMathSymbol{\zeta}{\mathalpha}{greek}{"10}
  \DeclareMathSymbol{\eta}{\mathalpha}{greek}{"11}
  \DeclareMathSymbol{\theta}{\mathalpha}{greek}{"12}
  \DeclareMathSymbol{\iota}{\mathalpha}{greek}{"13}
  \DeclareMathSymbol{\kappa}{\mathalpha}{greek}{"14}
  \DeclareMathSymbol{\lambda}{\mathalpha}{greek}{"15}
  \DeclareMathSymbol{\mu}{\mathalpha}{greek}{"16}
  \DeclareMathSymbol{\nu}{\mathalpha}{greek}{"17}
  \DeclareMathSymbol{\xi}{\mathalpha}{greek}{"18}
  \DeclareMathSymbol{\pi}{\mathalpha}{greek}{"19}
  \DeclareMathSymbol{\rho}{\mathalpha}{greek}{"1A}
  \DeclareMathSymbol{\sigma}{\mathalpha}{greek}{"1B}
  \DeclareMathSymbol{\tau}{\mathalpha}{greek}{"1C}
  \DeclareMathSymbol{\upsilon}{\mathalpha}{greek}{"1D}
  \DeclareMathSymbol{\phi}{\mathalpha}{greek}{"1E}
  \DeclareMathSymbol{\chi}{\mathalpha}{greek}{"1F}
  \DeclareMathSymbol{\psi}{\mathalpha}{greek}{"20}
  \DeclareMathSymbol{\omega}{\mathalpha}{greek}{"21}
  \DeclareMathSymbol{\varepsilon}{\mathalpha}{greek}{"22}
  \DeclareMathSymbol{\vartheta}{\mathalpha}{greek}{"23}
  \DeclareMathSymbol{\varpi}{\mathalpha}{greek}{"24}
  \DeclareMathSymbol{\varrho}{\mathalpha}{greek}{"25}
  \DeclareMathSymbol{\varsigma}{\mathalpha}{greek}{"26}
  \DeclareMathSymbol{\varphi}{\mathalpha}{greek}{"27}
  \DeclareSymbolFont{otone}{OT1}{cmr}{m}{n}
  \DeclareMathSymbol{\Gamma}{\mathalpha}{otone}{0}
  \DeclareMathSymbol{\Delta}{\mathalpha}{otone}{1}
  \DeclareMathSymbol{\Theta}{\mathalpha}{otone}{2}
  \DeclareMathSymbol{\Lambda}{\mathalpha}{otone}{3}
  \DeclareMathSymbol{\Xi}{\mathalpha}{otone}{4}
  \DeclareMathSymbol{\Pi}{\mathalpha}{otone}{5}
  \DeclareMathSymbol{\Sigma}{\mathalpha}{otone}{6}
  \DeclareMathSymbol{\Upsilon}{\mathalpha}{otone}{7}
  \DeclareMathSymbol{\Phi}{\mathalpha}{otone}{8}
  \DeclareMathSymbol{\Psi}{\mathalpha}{otone}{9}
  \DeclareMathSymbol{\Omega}{\mathalpha}{otone}{10}
  \DeclareSymbolFont{syms}{OML}{cmm}{m}{it}
  \DeclareMathSymbol{\partial}{\mathord}{syms}{"40}
  \DeclareMathAlphabet{\mathbold}{OML}{cmm}{b}{it}
  \DeclareSymbolFont{largesymbols}{OMX}{cmex}{m}{n}

}